# Physics Teachers' Perceptions about Diagnostic Assessment of Students' Physics Misconceptions: A Phenomenological Study

Mutmainna[1,2], Edi Istiyono[1], Haryanto[1],
Retnawati[1], and Caly Setiawan[1]
[1]Educational Research and Evaluation Doctoral Study Program, Universitas Negeri Yogyakarta, Indonesia
[2]Physics Education Study Program, Universitas Sulawesi Barat, Indonesia

Misconception is the difference between a student's idea and his scientific concept. Physics learning needs to be designed by teachers to minimize misconceptions. Different diagnostic tools have been developed and used by researchers to identify student misconceptions. This study addressed to uncover the experiences of eight physics teachers on how they apply diagnostic tools to detect misconceptions experienced by students investigated through phenomenological studies. Physics teachers are invited to volunteer to participate in this research. Physics teachers involved in this research have diversity based on the period of experience teaching physics, geographical area, and the level of students faced. A semi-structured interview (duration of 40 to 60 minutes) through virtually was conducted by the author for all physics teachers involved in this study. Personal Practice Theories (PPTs) by Cornett was chosen as a framework to reveal the experiences and perspectives of physics teachers in making judgments about students' misconceptions of Physics. The findings revealed that four themes were obtained. These themes are (1) personal experiences of Physics teachers, (2) Physics teachers' knowledge, (3) practices in assessing Physics misconceptions, and (4) teachers' reflections on future assessment practice.

When a teacher explains a new concept to students in the classroom, each student constructs their own knowledge uniquely, influenced by their individual life experiences (Allen, 2014). This indicates that students' knowledge constructions (mental models) are not created from scratch but are built upon pre-existing knowledge structures and are interconnected (Allen, 2014; Hammer, 1996; Smith III et al., 1994; Southerland et al., 2001). DiSessa called this the "knowledge in pieces view" (diSessa, 1988, 1993) and considers part of the student's natural development process (Gilbert & Watts, 1983). A learning approach that emphasizes how students construct their conceptual understanding is the constructivist approach (Resbiantoro & Setiani, 2022), introduced by Swiss psychologist Jean Piaget, known as the father of constructivism (Allen, 2014). Additionally, for a new fact or concept to be meaningful to students, it must fit into their previously constructed knowledge model. Otherwise, students are less likely to remember the new information later (Allen, 2014). Mental models derived from students' life experiences that differ from scientifically accepted knowledge are termed misconceptions (Allen, 2014; Driver & Easley, 1978; Martin, 1998;

Neidorf et al., 2020; Tippett, 2004). Misconceptions are also known by various terms such as preconceptions, alternative frameworks, intuitive beliefs, spontaneous reasoning, alternative conceptions, children's science, naïve beliefs, and mental models (Clement et al., 1989; diSessa, 1988; Greca & Moreira, 2002; Karpudewan et al., 2017; Klammer, 1998; McCloskey et al., 1980)

Identifying student misconceptions is generally an easy task for a teacher. However, the most challenging part is correcting students' understanding to transform their misconceptions into scientifically acceptable ideas (Allen, 2014). Research findings consistently show that misconceptions are deeply rooted and persist even after instruction (Allen, 2014; Eryilmaz, 2002; Geddis, 1951). Several previous studies have been conducted on instructional strategies proven effective in addressing misconceptions through various student activities such as discussions and experiments. Teachers play a crucial role in successfully implementing these strategies (Gomez-Zwiep, 2008). However, research reported by Gomez-Zwiep (2008) indicates that although teachers are largely aware of misconceptions, they do not understand how to develop their roles in instruction, and the findings show that there are similarities in misconceptions experienced by both teachers and students on certain topics (Burgoon et al., 2011). Meanwhile, Moodley & Gaigher (2019) found that teachers' understanding of misconceptions is influenced by their educational background. Therefore, these studies indicate that scientific research on misconceptions and the role of teachers in addressing them is urgently needed for continued investigation.

Fundamental scientific studies related to how students construct their knowledge, such as those linked to social interaction by Lev Vygotsky in the 1930s (Allen, 2014; Jenkins, 2000; Toh et al., 2003; Weil-Barais, 2001; Windschitl, 2002), and studies discussing learning strategies proven effective in addressing students' misconceptions (Ausubel et al., 1978; Posner et al., 1982). In the 21st century, some scientific studies include investigations into misconceptions among intern or novice teachers (L. Halim & Meerah, 2002), comparative case studies on the concept of prior knowledge among preservice, first-year, and expert teachers (Meyer, 2004), and investigations into misconceptions among teachers and students (Burgoon et al., 2011). These scientific studies generally emphasize the need to pay attention to the professionalism of teachers, particularly in reducing students' misconceptions. The study of misconceptions remains a continuously researched topic within the physics education research (PER) community (Docktor & Mestre, 2014). Recent examples of research include investigating students' misconceptions by examining the consistency of their answers through isomorphic questions (questions that measure the same ability and are developed in multiple items) (Michael, 2022); students' misconceptions using multiple-choice questions by applying network analysis (Scott & Schumayer, 2018a; Wells et al., 2019a, 2020a; Wheatley et al., 2022), and converting open-ended questions into multiple-choice questions to investigate students' misconceptions (Erceg, 2016). These studies highlight various efforts to delve into how and why misconceptions occur among students. For instance, studies on students' misconceptions using network analysis (Scott & Schumayer, 2018a; Wells et al., 2019a, 2020a; Wheatley et al., 2022) found that the response patterns (network modules) obtained are related to how students' misconceptions are held on certain topics and are interconnected.

Given these findings, it is deemed necessary for teachers to have the ability to diagnose students' misconceptions (Anam Ilyas, 2018). Kaltakci Gurel et al. (2015) compared various diagnostic instruments in science to assess students' misconceptions. They identified 273 articles published between 1980 and 2014 to reveal the strengths and weaknesses of each diagnostic tool. Subsequently, (Resbiantoro & Setiani, 2022) reviewed 72 international journal articles on diagnostic methods, causes, and remediation strategies for misconceptions published between 2005 and 2020. The results identified various diagnostic tools for physics misconceptions: interviews, open-ended tests, multiple-choice tests, and multiple-tier tests.

Knowledge of the causes and diagnostic abilities forms the basis for determining remediation strategies or preventing misunderstandings. Interviews are suitable for uncovering new misconceptions with a few participants, while four-tier tests are more effective for many participants. The most effective remediation strategies through a conceptual change approach are inquiry-based learning, simulation-based experiments, and conceptual change texts. The findings from these studies encompass diagnostic tools, remediation strategies, the origins of misconceptions, and unexplored physics topics, particularly focusing on the outcomes of qualitative research. (revised)

Referencing previous research findings, it is stated that teachers need to understand misconceptions in Physics, from the stages of diagnosing, identifying causes, to the remediation process (diSessa, 1993; Gomez-Zwiep, 2008; Leonard et al., 2014). Resbiantoro & Setiani (2022) also recommends that the development of diagnostic tools and remediation methods remains an important topic for future research, indicating the need for ongoing studies. This article is presented as a preliminary study, aiming to initiate an investigation into Physics teachers' perceptions of diagnostic assessments of students' physics misconceptions as a basis for future development studies. Furthermore, the research findings will enrich the learning community, contribute to the existing body of literature, and serve as a valuable reference for future Physics educators. This will be beneficial for curriculum developers. A phenomenological study framework has been chosen to achieve the investigative goals. This study is guided by the following research question: (revised)

RQ1 : How do physics teachers perceive students' misconceptions in physics?
RQ2 : What knowledge and methods of teachers employ to diagnose students' misconceptions in physics?
RQ3 : How do they plan future assessments to evaluate students' physics misconceptions?

## Literature Review

### A Review Misconceptions in Physics

The discussion of misconceptions is inseparable from the learning theories propagated by earlier scholars, particularly the proponents of constructivism, Jean Piaget and Lev Vygotsky (Harasim, 2017). One of Piaget's renowned views is that children think about the world in ways that are fundamentally different from adults (Smith III et al., 1994). Research in the field of Physics Education related to misconceptions often refers to the P-Prims perspective or Knowledge in Pieces introduced by diSessa (1993). According to (diSessa, 1993), the process of concept formation in students is linked to primitive knowledge and phenomenological knowledge, which manifest as fragmented pieces of knowledge (Knowledge in Pieces). These fragments of knowledge coalesce into cohesive, interconnected knowledge (Hammer, 1996). Some experts consider this stage as a natural developmental process in a student (Gilbert & Watts, 1983) and view it as forming, changing, and growing dynamically (Brown, 2014). Researchers also refer to misconceptions using other terms such as "naive beliefs" (McCloskey et al., 1980), "preconceptions" (Clement, 1982; Clement et al., 1989), "alternative conceptions" (Hewson & Posner, 1984; Taber, 2008), "naive theories" (Tobin, 2006), and "alternative frameworks" (Driver, 1983; Driver & Easley, 1978).

(Resbiantoro & Setiani, 2022) reported that most studies aim to identify or diagnose misconceptions in physics at the undergraduate level, among high school students, and pre-service teachers (Abrahams et al., 2015; Kanıawatı et al., 2019; Martínez-Borreguero et al., 2018; Moli et al., 2017). Misconceptions among undergraduates tend to be more complex, whereas for pre-service teachers, most researchers focus on this group because the misconceptions held by pre-service teachers have the potential to be transferred to students when they become teachers (Kumandaş et al., 2019). Besides this objective, studies generally explore the impact of treatments given to address students' misconceptions (Korganci et al., 2015; Trundle et al., 2007), investigate the causes of misconceptions (Erman, 2017; A. Halim et al., 2019), examine cognitive structures (H. Kang et al., 2010), and develop diagnostic tests for physics misconceptions (Desstya et al., 2019).

Furthermore, (Kaltakci Gurel et al., 2015) reported his research findings by detailing various types of instruments used to detect students' misconceptions, including the Force Concept Inventory (FCI) (Hestenes et al., 1992), Force & Motion Conceptual Evaluation (FMCE) (Thornton & Sokoloff, 1998), Mechanics Baseline Test (MBT) (Hestenes & Wells, 1992), Energy & Momentum Conceptual Survey (EMCS) (Singh & Rosengrant, 2003), Test of Understanding Graphs in Kinematics (TUG-K) (Beichner, 1994), Electric Circuits Conceptual Evaluation (ECCE) (Sokoloff, 1996), Conceptual Survey in Electricity & Magnetism (CSEM) (Maloney et al., 2001), Brief Electricity & Magnetism Assessment Tool (BEMA) (Ding et al., 2006), Light and Spectroscopy Concept Inventory (LSCI) (Bardar et al., 2006), Quantum Mechanical Visualization Inventory (QMVI) (Cataloglu & Robinett, 2002), and Mechanical Waves Conceptual Survey (MWCS) (Tongchai et al., 2009). These instruments are openly accessible (American Association of Physics Teachers). Thus, from the explanations provided, it is evident that student misconceptions are investigated through the development of inventory tests.

Suprapto, N. (2020) identifies four sources of misconceptions: (1) students, where misconceptions can stem from their everyday experiences or deeply rooted prior learning processes (Allen, 2014; Neidorf et al., 2020), or other factors such as humanistic thinking, incomplete or incorrect reasoning, false intuition, stages of cognitive development, students' abilities, and learning interest; (2) teachers, where misconceptions may be influenced by their educational background (Moodley & Gaigher, 2019); (3) instructional materials and literature; and (4) teaching methods.

## Diagnostic Tools of Students' Physics Misconceptions

Various instruments for diagnosing misconceptions have been developed and utilized by researchers (Resbiantoro & Setiani, 2022). Generally, tools for diagnosing misconceptions in the field of Physics Education (Kaltakci Gurel et al., 2015; Resbiantoro & Setiani, 2022) include interviews and open-ended tests, multiple-choice tests, and multi-tier tests (such as two-tier tests, three-tier tests, four-tier tests, etc.). Each type of test has its own advantages and disadvantages (Kaltakci Gurel et al., 2015; Resbiantoro & Setiani, 2022). The following section will provide a brief explanation of the characteristics of these diagnostic tools for misconceptions.

Open-ended tests are characterized by their ability to capture students' ideas through the responses they provide (Mintzes et al., 2005). The advantage of open-ended tests is that they allow students to express their understanding in their own words. Additionally, with open-ended tests, students often provide unexpected answers (Kaltakci Gurel et al., 2015; Mintzes et al., 2005). However, the disadvantages of open-ended tests include: (1) requiring considerable time to analyze students' answers; (2) some students tend to give brief responses, which can hinder researchers from diagnosing misconceptions; (3) scoring remains relatively problematic in open-ended tests (Kaltakci Gurel et al., 2015).

The next diagnostic tool for identifying misconceptions is the multiple-choice test. (Fisher & Frey, 2014) describe that a multiple-choice test generally consists of two parts: the stem (question body) and the list of answer choices. In the list of answer choices, only one option is the most correct, while the others serve as distractors, which are incorrect answer options. Students with misconceptions are detected by the questions they answer incorrectly, being misled by the distractor options. In Physics Education, there are several types of multiple-choice instruments that have been used in various studies to diagnose students' misconceptions (Brewe et al., 2016; Scott & Schumayer, 2018b; Stewart et al., 2021; Wells et al., 2019b, 2020a, 2020b). The advantages of multiple-choice tests in diagnosing misconceptions include simpler test administration, quick scoring, objective scoring, and ease of application even to large sample size (Kaltakci Gurel et al., 2015). However, there are also challenges in using multiple-choice tests, such as not providing students with the opportunity to give in-depth responses, the tendency for students to guess answers, and the longer time required to construct the questions and answer choices (Brookhart & Nitko, 2019; Kaltakci Gurel et al., 2015; Mintzes et al., 2005).

Another diagnostic tool is the Multiple-Tier Tests (MTT), which include two-tier tests (2TT), three-tier tests (3TT), and four-tier tests (4TT) (Resbiantoro & Setiani, 2022). Essentially, the structure of Multiple-Tier Tests (MTT) is like multiple-choice tests. The difference lies in the additional questions that follow the selection of an answer choice, which then confirm the reason for choosing that option (two-tier tests (2TT) (Barniol & Zavala, 2016; A. Halim et al., 2018; Ivanjek et al., 2021), select an answer choice, confirm the confidence level in the given answer, and provide a reason for choosing the option (3TT) (Caleon & Subramaniam, 2010; Rusilowati et al., 2021), and also select an answer choice, confirm the confidence level in the given answer, provide a reason for choosing the option, and confirm the confidence level in the reason provided (4TT) (Kaltakci-Gurel et al., 2017; Tumanggor et al., 2020), and potentially more. The strengths and weaknesses of multiple-tier tests are generally like multiple-choice tests (Kaltakci Gurel et al., 2015). Some additional advantages of multiple-tier tests include their better capability to capture students' misconceptions compared to regular multiple-choice tests. However, their weaknesses generally include the tendency to underestimate the extent of knowledge gaps because they cannot determine if the respondent is confident in their answer, overestimate students' scores, and the longer time required for students to answer the questions, which can lead to student fatigue (Kaltakci Gurel et al., 2015).

In addition to tests (open-ended, multiple-choice, and multi-tier tests), diagnosing student misconceptions can also be conducted through interviews (Abrahams et al., 2015). Interviews are a method that can detect student misconceptions in more detail (Brookhart & Nitko, 2019; Hestenes & Wells, 1992). Some advantages of using interviews to diagnose misconceptions include the ability to gather in-depth information from students and the flexibility of questions to adapt to each student's response. However, Kaltakci Gurel et al. (2015) outlines several challenges in using interviews as a tool for diagnosing misconceptions, including (1) interviews require a significant amount of time, especially when involving large samples, (2) teachers need interview skills, including building trust with students to facilitate in-depth interviews, and (3) there is a tendency for subjectivity in evaluating interview results.

**Personal Practice Theories (PPTs)**

In the context of this study, Personal Practice Theories (PPTs) by Cornett (Cornett, 1990)) were chosen as the framework to explain the phenomenon of teachers' experiences in conducting diagnostic assessments of students' physics misconceptions. The fundamental concept explained in Cornett's theory is how teachers' PPTs are influenced by the interaction between teachers' prior life experiences (Personal), teachers' thematic set of beliefs (Theories), and teachers' classroom experiences in assessing physics misconceptions (Practice). Besides Cornett (Cornett, 1990), several previous researchers have applied PPTs, such as discussing practical knowledge in teaching and the development of practical teaching theories (Sanders & McCutcheon, 1986), the landscape of teachers' professional knowledge(Clandinin & Connelly, 1996), exploring science teachers' epistemological beliefs regarding the use of laboratory activities (N.-H. Kang & Wallace, 2005), and investigating the impact of short-term study programs for in-service teachers (He et al., 2017). In addition, Cornett's theory is also guided by Personal Practice Assessment Theories (PPATs) introduced by Box et al. (Box et al., 2015). PPATs are an extension of Cornett's theory. This theory was chosen because of its relevance to assessment. Among the two theories, we use Cornett's theory as the main theoretical framework. This decision was made considering the research questions we aim to explore.

The core of our study is to explore and understand teachers' perceptions of diagnostic tests for physics misconceptions, their experiences in diagnosing students' physics misconceptions, whether they possess adequate knowledge about the development of diagnostic tools for students' physics misconceptions, and whether teachers have established best practices for diagnosing students' physics misconceptions. Both theories (Box et al., 2015; Cornett, 1990) and these questions guided us in developing the interview protocol. The presentation of science lessons, including Physics, from elementary to higher education levels, showed similar issues related to physics misconceptions (Suprapto, 2020). Considering the sources of misconceptions (students, instructional materials/literature, teaching method, and teacher) (Suprapto, 2020), teachers play a central role in minimizing physics misconceptions. Teachers act as learning facilitators for students (Brookhart & Nitko, 2019), are responsible for selecting credible instructional materials/literature for teaching, and when teachers understand physics concepts, they can choose appropriate teaching methods (Carpendale & Cooper, 2021).

The application of PPTs in this study can explain teachers' prior life experiences (Personal) are believed to influence their positioning in front of their students. This can be observed, for instance, in interactions with students who have strong convictions about their understanding of concepts they believe to be correct. Furthermore, a set of interconnected beliefs (Theories) is also believed to impact PPTs, including in diagnosing students' misconceptions. This can include educational background, teaching experience, and so on. Ultimately, teachers' personal experiences and theories will guide their practice in diagnosing students' physics misconceptions in the classroom. For example, guiding teachers in deciding on the chosen teaching strategies, the media used, where the learning takes place, and so on. From the practices carried out by teachers, experiences will be generated that will influence the personal and theoretical components of the teachers.

To clarify our positions as authors regarding the research topic under investigation, we outline the profile of each author. The first author is a Ph.D. student in Educational Research and Evaluation at Yogyakarta State University since January 2022. The first author has been a lecturer in Physics Education at the Faculty of Teacher Training and Education, West Sulawesi University, Indonesia, since 2019. The first author's educational background includes a bachelor's and master's degree in physics education. Currently, the first author is focusing on measurement concentration in the Ph.D. program at Educational Research and Evaluation at Yogyakarta State University. The chosen dissertation study relates to diagnostic tests for physics misconceptions. The second author has been a lecturer in the Ph.D. program in Educational Research and Evaluation at Yogyakarta State University since 2015. The second author's expertise is in Measurement, Assessment, and Evaluation of Physics Education. Several of the second author's research projects align with the research topic, particularly concerning diagnostic tests for physics misconceptions, which have been explored in recent years. The third author has been a lecturer in the Ph.D. program in Educational Research and Evaluation at Yogyakarta State University since 2010. The third author's expertise is in Research and Evaluation of Electrical Engineering Education. The third author teaches courses relevant to the research topic and supervises Ph.D. students in the measurement concentration. The fourth author has been a lecturer in the Ph.D. program in Educational Research and Evaluation at Yogyakarta State University since 2009. The fourth author's expertise is in Mathematics Education Assessment. The fourth author teaches courses relevant to the research topic and supervises Ph.D. students in the measurement concentration. Finally, the fifth author has been a lecturer in the Ph.D. program in Educational Research and Evaluation at Yogyakarta State University since 2017. The fifth author's expertise is in Physical Education and Sports Pedagogy. The fifth author teaches courses relevant to the research topic, particularly in Qualitative Research Methods and Qualitative Data Analysis Techniques.

## Methods

### Study Design

The approach used in this study is the phenomenological approach. The phenomenological study is considered an appropriate research design to explore the experiences of Physics teachers in making assessments about their students (Santoso et al., 2022). Several expert views in Moustakas (1994) guided us in processing the obtained data. Phenomenology is an approach that involves returning to experiences to obtain a comprehensive description (Van Kaam, 1967) that describes what is felt, sensed, and known by consciousness (Kockelmans, 1987) in the form of naive and general descriptions (Giorgi, 1985) to reveal the structure of meaning and how that meaning is created (Von Eckartsberg, 1986). Similarly, (Creswell & Poth (2016) explains that phenomenology is a qualitative approach that aims to find the essence of a person's experience regarding a particular phenomenon. From these various explanations, the authors are guided to define phenomenology in this study as an approach aimed at uncovering the naive experiences of Physics teachers felt directly, by attempting to analyze the structure of the meaning of experience and how that meaning is created, thus discovering the essence of the experiences they undergo, as framed by the theories of Cornett (1990) and Box et al., 2015).

### Participants

Eight physics teachers at secondary schools and higher education (five males and three females), were studied in this research. The study participants reside in three regions of Indonesia, with two physics teachers from the eastern region, four teachers from the central region, and two teachers from the western region of Indonesia. The complete demographic conditions of the participants are presented in Table 1. Participants for this study were recruited voluntarily by the authors via an online form disseminated through social media (WhatsApp). In selecting participants, we were guided by the following criteria:

1. Participants confirmed their willingness to participate via the online form distributed by the researchers. For those who agreed to participate, the form requested a contact telephone number for further communication with the researchers.
2. Participants must be physics instructors (at secondary schools or higher education).

For participants who met both criteria, the researchers subsequently sent a private message to the participants via WhatsApp to schedule an interview. The private message sent to participants contained an introduction of the researcher and information deemed essential for the participants to know in advance. This information included the interview topic, estimated duration of the interview, the conduct of the interview (in this study being conducted via Zoom Meeting), and technical details (such as the sharing of virtual interview links). Fortunately, eight participants confirmed their availability shortly after the message was sent. The study employed semi-structured interviews lasting approximately 45 - 60 minutes. During the interview, the researcher provided an initial explanation and sought permission from the participants to record the interview process. Recording commenced once participants had given their consent. As part of the researchers' obligation to protect all collected participant data, we were guided by Lahman et al. 2011). An example of the application of ethical clearance in this study includes the participants' names listed in Table 1, which have been anonymized arbitrarily to maintain and respect participants' privacy.

**Table 1**
*Demographic Conditions of Participants*

| Participan Name (Anonim) | Age (Year) | Teaching Experience (Years) | Educational Qualifications | Level of Education Taught |
|---|---|---|---|---|
| Baim | 30 | 5 | Master of Physics Education | Higher Education |
| Mita | 33 | 8 | Master of Physics Education | Higher Education |
| Aswar | 32 | 7 | Bachelor of Physics Education | Senior High School |
| Dodi | 37 | 12 | Bachelor of Physics Education | Senior High School |
| Dani | 30 | 4 | Bachelor of Physics Education | Senior High School |
| Hardi | 33 | 8 | Bachelor of Physics Education | Senior High School |
| Budi | 45 | 20 | Master of Physics Education | Senior High School |
| Nana | 33 | 8 | Bachelor of Physics Education | Senior High School |

## Data Collection

Interviews were chosen as the data collection method for this study. Interviews allow researchers to delve into areas of subjective attitudes and personal experiences (Peräkylä & Ruusuvuori, 2008) and are considered one of the critical sources of information (Yin, 2009). The first author acted as the primary interviewer in the data collection process. Interviews were scheduled by the researcher, one participant per day, to maintain consistency in understanding each teacher's experiences. During the interviews, the researcher employed semi-structured interviews and recorded the conversations using a virtual meeting application (Zoom). As soon as an interview was completed, we directly transcribed the recent conversation to ensure that the data obtained was well-documented and to build a deeper understanding of the physics teachers' experiences under investigation. Another effort made by the researchers to obtain optimal data from participants was to ask for their willingness to be re-contacted if further confirmation was needed for statements made in the initial interview. Out of our eight participants, two were re-contacted for further clarification on several points from the previous interviews.

**Table 2**

*Sample Questions on the Interview Protocol*

| No. | Theoretical Domain | Question Example |
|---|---|---|
| 1 | **Teachers' Personal** Exploration the personal experience of the Physics teacher | What is the character of your students in learning Physics? What are some of your challenges in teaching? |
| 2 | **Teachers' Theories** Seeking personal experience with the theory of assessment of Physics teachers | What methods can be used to diagnose students' misconceptions in Physics so far? Have you ever received any training related to diagnostic assessment of Physics misconceptions? |
| 3 | **Teachers' Practice** Seeking meaningful teacher practices and assessment development plans. | How do you diagnose your students' misconceptions in Physics? Have you had any meaningful experiences while diagnosing your students' misconceptions in Physics? What are your future plans? |

## Data Analysis

The recorded conversation data is consistently maintained in each interview as an audio file. However, the transcription engines in Indonesian and corresponding regional languages are currently limited. Our participants speak at least two languages, both Indonesian and their ethnic languages. Possible drawbacks can be introduced if we rely solely on transcription machines. Therefore, we decided to transcribe it manually. Our transcribes are retained in text format (.txt). Content analysis is then performed through RQDA (R-based Qualitative Data Analysis) packages in the R language to assist in the coding management of each transcript segment. Since RQDA is not available in the latest version of the R Program, we use RQDA by downloading via GitHub Developer developed by R Core Theme (2022).

Stages the data analysis step begins with identifying the attributes of research participants. We classify participants based on several criteria including teaching experience, geographic region, and the level of students they are dealing with. The second step is to do an open coding or known as an inductive coding (Setiawan, 2022). This is done by reading and interpreting repeatedly each part of the participant's experience. Third, the collection of coding that has been marked is then interpreted in depth and repeatedly to be sorted into categories based on the experience of physics teachers in assessing students' physics misconceptions and limitations in conducting assessments carried out so far. We subsequently analyzed the categorized data using a thematic approach (Braun & Clarke, 2022). In this step we produce a codebook as a research base to help answer our research questions. Another step taken to ensure that the data we have recorded is in accordance with the experience of the participants is to do a member check with the participants. This step is done by returning interview transcribe to participants and asking them to verify the statements listed and asked to comment if there are statements that are considered necessary for redaction to be corrected. Likewise, the results of the analysis and draft of the publication manuscript were also consulted with the participants. This step is taken as part of maintaining the concept of trust as internal and external validity in the research we conduct. Thus, it can be said that our findings have represented exactly the essence of the teacher's experience in assessing students' physics misconceptions and the experience they gained during learning, especially in conducting assessments.

## Limitations

Researchers must recognize the limitations of their studies. The phenomenological analysis we conducted with eight Physics teachers, exploring their perceptions of diagnostic tests for misconceptions, involved voluntary participants from three regions of Indonesia: the Eastern, Central, and Western parts. Although our aim is not to generalize our findings to other contexts, we want to confirm that the results reported in this study solely reflect the data obtained from participants who voluntarily took part. Consequently, we may not have captured the full spectrum of Physics teachers' perspectives on their perceptions of Diagnostic Assessment of Students' Physics Misconceptions. Investigating how teachers adapt to students based on their generational characteristics (e.g., Generation Z and Generation Alpha), which has not been explored in this study, could be a project for future research.

# Findings

The research findings are generally presented by referencing both theoretical frameworks, namely Cornett's (1990) theory and Box et al.'s (2015) theory, to address each research question. Based on the thematic construction from the analyzed data (coded and categorized into themes), four themes were identified. These themes are (1) personal experiences of Physics teachers, (2) Physics teachers' knowledge, (3) practices in assessing Physics misconceptions, and (4) teachers' reflections on future assessment practice. Below is a detailed explanation of these themes. As an additional clarification, in presenting the research findings, the term "physics teacher" is understood to refer to individuals teaching Physics, whether at the secondary school level or in higher education institutions.

## 1. Personal experiences of Physics teachers

### 1.1 Teachers’ motivation

We began the interviews by asking participants to confirm their general identities as listed in Table 1. PPTs depict the complex interactions between beliefs, knowledge, and practice of teachers, including those based on personal experiences (such as their lives as students and parents) (Box, 2015; Cornett, 1990). This prompted us to start the interviews by asking what motivated them to become Physics teachers. In Cornett's theory, this relates to teachers' prior life experiences (personal).

There are various personal reasons confirmed by the teachers for choosing Physics teaching as a profession. The first reason was inspired by the teaching methods of their **Physics teachers when they were students**, as illustrated by the following statement:

> ...my motivation to study Physics began in high school. It was because my Physics teacher's teaching method was excellent, which made me understand Physics faster than other subjects. (Aswar)

> ... so initially, I chose the Physics major due to encouragement from my Physics teacher. In high school, I was close to him, so he suggested I take Physics (Nana)

> ... in elementary school, I remember how my teachers supported me, how they lovely welcomed me at the door (Budi)

Similarly, Hardi's reason for becoming a Physics teacher is in line with this. He stated, “almost everyone experiences that the subject is not related to the subject itself but indirectly to the teacher. At that time, my Physics teacher made explanations very enjoyable”. From these quotes, we gathered that teachers, as role models, are one of the reasons someone chooses to become a Physics teacher and an inspiration in their profession, as stated by Aswar, “I consider that I must be better than the teacher who taught me.”

The second reason is an **affinity for calculations and science subjects**. This is evident in the statement from Hardi which said “I find Physics very interesting. Besides requiring reasoning, Physics must be related to real-life contexts”. Another expression from Budi states, “since high school, I have enjoyed calculations. I enjoy Mathematics, Physics, Chemistry....” This statement aligns with Nana's, “I like Physics. I really like it. I enjoy calculations. So, when my Physics teacher suggested it, I finally chose the Physics major.” From Nana and Budi's statements, it is evident that besides the teacher's role in providing **encouragement and motivation**, their fondness for mathematics and science subjects also plays a part.

Another perspective on the reason for choosing Physics teaching as a profession was expressed by Budi “I saw education as the **quickest route to getting a job**. That's why I became a teacher.” Meanwhile, Aswar expressed his **desire to advance society**, as seen in his statement, “I observed that the community here, especially the students I teach, have significant technological backwardness.”

Based on the above explanations, the teachers have diverse motivations for becoming Physics teachers. The participants' confirmations align with Kyriacou et al.'s (2000) explanation of the three fundamental reasons individuals choose the teaching profession. First, altruistic reasons, which see teaching as something beneficial to social life. Second, intrinsic reasons, which include the activity of teaching children and a person's interest in utilizing their knowledge. Third, extrinsic reasons, which encompass salary, social status, and so on.

### 1.2 Teachers' Perception of Their Profession

After interviewing the participants about what motivated them to choose the profession of Physics teaching, we then investigated whether this motivation influenced their perception of their status as Physics teachers. The teachers' perceptions of their profession were expressed in various roles. First, viewing themselves through the **lens of their experiences as students**. An example statement showing participants recalling their time as students is Aswar's remark “when facing students at school, I remember my days as a student.”. This is aligning with Hardi's statement, “my Physics teacher made explanations very enjoyable... Physics was easy to learn because it was very simple and closely related to everyday life.” Another statement was made by Nana, “although the teaching method was conventional, my teacher was excellent at explaining. That's when I fell in love with how he explained things.” Nana's admiration for her Physics teacher's way of explaining the material inspired her to practice the good methods her teacher had used, as indicated by her statement, “what I learned from my Physics teacher was to train competitiveness.”

Besides recalling their experiences as students, the teachers also positioned themselves **as friends** to their students, facilitators, and even parents. Aswar stated that a teacher should not be too close nor too distant from the students. Acting as a friend to the students can be achieved by sharing experiences, which Aswar believes can be valuable to the students, as he mentioned,

> ...if we can act as their peer, as a relative, sharing our experiences. So, the experiences I had as a teenager, because I teach high school, we share with the students. This helps them to discern and choose how they should behave during their teenage years. (Aswar)

Hardi, in addition to positioning himself as a friend to his students, also sees himself **as a manager/facilitator**. This is confirmed through his statement, “I now position myself firstly as a friend, and secondly as a manager.” Further, Hardi explains, “As a manager, I see myself as a facilitator for my students. I guide them on how to learn safely, comfortably, and most importantly, how to easily understand the material I teach.” Lastly, the teachers' perception of their profession also sees themselves **as parents** to their students. Aswar, for instance, states, 'There are moments when we see students facing problems, for example with their environment, and in such cases, we as teachers must step in as their surrogate parents at school.'

Based on the above descriptions, the data we obtained generally lead us to argue that teachers’ perceptions of their role as educators are influenced by their perspectives, considering their experiences both as students and as teachers. This aligns with the view that PPATs describe the complex interactions between teachers' beliefs, knowledge, and practices, grounded in personal experiences (such as their lives as students and parents) (Box, 2015; Cornett, 1990).

### 1.3 Teachers' Knowledge of Student Characteristics

As a further exploration of Cornett's (1990) and Box et al.'s (2015) theories, we sought to delve into the participants' knowledge regarding student characteristics. In this subtheme, we generally identified four key topics related to teachers' understanding of their students' traits, categorized into (1) student characteristics based on content delivery, (2) student characteristics during assessment, (3) student characteristics based on gender, and (4) student characteristics based on interest. These four categories are outlined as follows.

Teachers’ knowledge of **students’ characteristics based on content delivery** includes understanding when, where, and how to present the material effectively. For instance, Aswar mentioned that his students were more engaged and actively participated in lessons conducted outside the classroom. This is evident in his statement:

> ...so, the students prefer learning outside the classroom. I present the material in a simple manner, but the students can engage with it through activities. So, during the learning process, students are active. They play with water, play with trees, which is far more interesting for them than merely explaining concepts from textbooks in the classroom (Aswar).

To reinforce this point, Aswar shared a memorable experience teaching Physics, specifically on the topic of Archimedes' Law, as illustrated in the following excerpt:

> I once taught an 11th-grade class in 2014. I was teaching Archimedes' Law. When I first explained the theory, the students seemed confused. So, I took them outside. In front of the school, there was a river. I brought them there to play with water. I asked them to try lifting one of their friends. They found their friend heavy. But when I asked them to carry the same friend in the water, they felt their friend was lighter. That’s what we call buoyancy in fluids. That’s how I explained the material. (Aswar)

From this account, Aswar recognized that not only do his students enjoy being taught outside the classroom, but they also grasp the material more easily when it is presented with contextual examples. Moving on to statements from other participants, still related to student characteristics based on content delivery, there is confirmation of differences between students taught online and offline learning. Mita elaborated on her experiences with students before, during, and after the Covid-19 pandemic. The pre-pandemic period occurred in 2018. The pandemic stretched from late 2019 until mid-2023. The post-pandemic period followed mid-2023. Students before and after the pandemic were taught offline learning. During the pandemic, they were taught online learning. Mita stated:

> ...during the Covid-19 pandemic, lectures were conducted online. This condition impacted students' mastery of concepts. But in 2023, we are focusing on offline learning again. So, students' conceptual mastery is no longer the same as before (compared to students taught during the Covid-19 pandemic). Students in 2023 share similarities with those from 2018. (Mita)

Similarly, Dodi shared the same perspective and said, "The impact of Covid-19 was very palpable. The difference is very noticeable in the new high school students compared to those before the pandemic." Still related to the first category, another confirmation came from Nana, who noted that her students were more enthusiastic about learning when rewarded for participating in class. This is evident in her statement: "I usually prepare gifts from home and keep them in my pocket to give to the students who can answer questions. The students are motivated to learn when I do this."

The second category, **students' characteristics during assessment**, is represented by Dodi's statement: "...there are not many students who can answer questions correctly. Most students, when given multiple-choice questions, just guess by counting the buttons on their shirt." This aligns with Nana's observation: "When students take multiple-choice tests, they're done in three minutes." The third category, **students' characteristics based on gender**, is represented by Baim's statement: "In a class, there are usually only three or four male students. However, despite this imbalance between male and female students, the males tend to be more vocal in expressing their opinions." The final category, **students' characteristics based on interest** to Physics, is highlighted by Dodi's comment: "There are very few students who are genuinely interested," and Budi's statement: "I don't want to make Physics seem too intimidating for the students by overwhelming them with too many formulas."

These responses illustrate that the teachers know their students' characteristics in various contexts. Particularly regarding assessment, some teachers consider multiple-choice questions an instrument that might not align well with their students' traits. Additionally, the data gathered reveals how the participants attempt to take on their roles optimally by recognizing their students' characteristics, allowing them to position themselves wisely in content delivery and assessment.

Our interview began by asking participants to confirm their general identity as stated in 1. There are some basic questions before asking informants to share their views on diagnostic tests of misconceptions. These questions include what motivates them to become physics teachers, how they position themselves in the classroom and how they recognize the character of their students.

## 2. Physics Teachers' Knowledge

### 2.1 Perceptions of Physics Misconceptions (RQ1)

Based on the data analysis that we conducted; participants' perceptions of Physics misconceptions are categorized into two distinct groups. The first category pertains to teachers' perceptions of students' misconceptions in Physics, while the second category relates to teachers' views on the impact of these misconceptions. Each category is explained as follows.

Participants generally regard **students' misconceptions of Physics as a natural occurrence**, believing it to be rooted in their everyday experiences. This view is reflected in Aswar's statement "misconceptions at the beginning of learning Physics are natural due to the complexity of the concepts being taught." Aswar further elaborated: "Students come to class with prior knowledge and intuitions based on their daily experiences, which sometimes conflict with the actual principles of Physics."

Another response from Budi suggested that misconceptions are a part of students' incomplete or imperfect understanding. This is evident in his statement:

> ...it's not just about building [understanding] and identifying misconceptions, but also recognizing concepts that may not be fully grasped. The student isn't necessarily wrong, they just haven't fully understood yet (Budi).

Budi further emphasized that misconceptions in Physics are particularly difficult to change, especially when they stem from students' everyday lives. This is reflected in his statement: "They (students) find it very difficult to accept [the teacher's explanations]. In their daily life, they struggle to move away from everyday terms."

The second category pertains to teachers' perceptions of the **impact of Physics misconceptions**. In this section, we discovered not only teachers' perceptions of the impact of misconceptions on students, but also on prospective teachers. Several questions confirmed these aspects, as illustrated by the statements from Dani and Budi:

> ...Physics concepts are continuous. For instance, the concept of force was introduced in grade 10 and revisited in grade 11. So, if the concept of Newton's laws is misunderstood early on, the misconception will persist through grade 11 and 12, affecting later understanding. (Dani)

> ... [the student] will not be able to accurately analyze the subsequent material because a misconception can be fatal. If the foundation is wrong, the structure built upon it will be flawed and may collapse. The understanding formed won't be what was intended. There might be something built, but it won't be the correct structure. (Budi)

Regarding the impact of misconceptions on prospective Physics teachers, Budi noted "...and if they become Physics teachers, it could be dangerous because they might teach incorrect concepts" (Budi).

From the explanations above, teachers' perceptions of students' misconceptions in Physics are seen as natural and may arise from various conditions, one of which is students' everyday experiences. As a result, misconceptions can be difficult to change once they have taken root. This aligns with the views of Allen (2014) and Eryilmaz (2002). Moreover, Physics misconceptions can affect students' understanding of interconnected topics, such as the application of vector quantities in Force and Motion. Failure to comprehend vectors may hinder students when asked to determine the net force acting on an object, for example.

The impact of misconceptions on prospective teachers can lead to the propagation of incorrect concepts when teaching Physics and result in flawed assessments of students' answers. This resonates with the idea that one source of misconceptions is the teacher (Suprapto N, 2020; Meyer, 2004; Burgoon, 2011).

### 2.2 Diagnostic Tools of Misconception in Physics (RQ2)

In this section, we found notable differences in the knowledge of Physics teachers at the secondary school level compared to those in higher education. At the secondary level, some teachers confirmed that they had applied certain types of instruments discussed during the interviews, although they were unfamiliar with the specific terms for these instruments. Meanwhile, Physics instructors in higher education recognized various diagnostic tools for Physics misconceptions but had never implemented them with their students. An example excerpt reflecting participants' knowledge of diagnostic tools for Physics misconceptions is as follows:

> ...the four-tier diagnostic tool really delves deeper into students' answers. I agree with that. I've had training in developing four-tier diagnostic tests, but I haven’t yet applied such questions in my class. (Mita)

An example of a secondary school teacher's statement is reflected in Dodi's comment "...I haven't heard that term before, but I have used questions like that. I just didn't know the name for it" (Dodi). Based on the data obtained, it appears that **knowledge of diagnostic tools for Physics misconception is still limited among most teachers**, **except for those in higher education.**

## 3. Practices in Assessing Physics Misconceptions

### 3.1 Contextual element

In the research reported by Box et al. (2015), one aspect discussed, believed to influence the assessment practices of teachers, is contextual elements such as high stakes testing, administration, cultural norms of the school, and other factors. The data reported by Box et al. in their research present contextual elements in two categories: Internally constructed contextual elements and External contextual elements. Each category includes what has been facilitated by teachers and what remains as challenges.

In the semi-structured interviews, we conducted regarding contextual elements, we found several insights from participants pointing towards strengthening institutions/communities as learning resources for teachers and the curriculum that has been implemented. An exploration of strengthening institutions/communities as learning resources for teachers confirmed the training and activities provided by their communities as learning tools related to diagnostic tools for students' physics misconceptions. Meanwhile, the exploration of the curriculum confirmed participants' perceptions of the existing curriculum. We investigated whether the curriculum influences how they view physics misconceptions and how they assess these students’ physics misconception.

The first part, **strengthening institutions/communities** as learning resources for teachers, yielded two categories: training for middle school students' physics misconceptions and training for higher education students' physics misconceptions. For training on middle school students' physics misconceptions, six participants confirmed that such training had never been conducted. This is evident in Dodi's statement, 'as for the understanding of concepts, there hasn't been any.' Similarly, Nana confirmed, 'in the teacher mover [one of the teacher communities], there was no discussion on assessment.” The training participants have received generally revolves around curriculum socialization with each curriculum change. This was confirmed in Aswar's statement, “...if the training is directly related to Physics, there have been several sessions from the central government, but the topic was on strengthening the concepts of the 2013 curriculum (Aswar).”

An additional insight from this exploration is the limited functionality of subject-specific teacher communities. One hindering factor is the inadequate transportation and limited internet connectivity in certain areas. This is evident in Nana's statement, 'in our place, there are still islands with no network at all.' Another confirmation from Aswar states, 'regarding the physics teacher community here, in our place, communication is not intensive because the schools are very far apart (Aswar).'

For the higher education level, training on physics misconceptions was indicated by Mita, who mentioned providing training to a teacher community on the Four-Tier Test: 'I provided training materials on creating four-tier questions.' Community strengthening in higher education was confirmed through Mita's statement, 'so in one semester, every month, there are small seminars where a lecturer provides material.'

The second part focuses on the **curriculum**. Some participants, particularly from secondary schools, shared their experiences with the curriculum when national exams were still implemented and their relation to physics misconceptions. Budi confirmed, 'in previous curricula, there was indeed a high chance of misconceptions.' Budi further elaborated, 'because our target was the national exam. It seemed like how diligently we memorized formulas, definitions, and practiced problems, not about understanding the concepts.' Another confirmation about the curriculum after the national exams were no longer implemented came from Dodi, 'in the Merdeka curriculum, even in Grade X, we have a lot of time [to learn], so we can start learning with the students without too many formulas.'

From this section, we gathered that the content density in the curriculum during the national exams influenced how teachers presented material in class to meet the exam targets, as Budi said, 'no matter what, the students had to be able to solve the problems, not physics understanding.' The data obtained indicate that both community strengthening and the current curriculum influence participants' perceptions in performing their duties as Physics teachers, especially regarding students' Physics misconceptions.

### 3.2 Assessment of Physics Misconceptions Teachers' Practice (RQ2)

Misconceptions in Physics can arise due to incorrect interpretations of fundamental Physics concepts. Therefore, Physics teachers need to conduct effective assessments to identify and address these misconceptions among their students. Our investigation into the practices of assessing Physics misconceptions among participants revealed two categories. The first category is assessment practices in the form of oral questions/interviews/direct questioning, and the second category is assessment practices in the form of tests.

Assessment practices in the form of **oral questions/interviews/direct questioning** were chosen by some participants as the most effective method for several reasons. First, it does not require extensive technical preparation. Second, it can provide quick feedback. Third, when diagnosing misconceptions through tests, participants often worry that the confirmed answers are a result of collaboration with friends. Some confirmations supporting this include Budi's statement "...I do it through direct discussions with the students so that I can provide immediate feedback to them." This is consistent with statements made by Nana and Dodi

> With interviews, we can immediately identify students' misconceptions. When given questions to answer, they sometimes collaborate with friends, making it difficult for us as teachers to detect the actual data/conditions (Nana).

> ...to anticipate misconceptions, I often walk around during class. So, I asked them [students] to explain what they have written to me (Dodi).

In addition to using interviews, we also confirmed from participants that diagnosing misconceptions is not conducted as a separate single activity but rather **integrated into formative tests**. This is evident in Dani's statement,

> When it comes to concepts, I usually use oral tests. Oral tests are more effective in my opinion because as we teach, we can see if the students understand or not. Secondly, through formative tests, we can gauge the extent of their understanding (Dani).

Mita's statements are more complex compared to other participants. The confirmation provided explains which aspects of teaching allow for diagnosing misconceptions. This is confirmed through the following statements:

> The method to diagnose misconceptions usually begins with examples or phenomena occurring in the students' environment. Typically, these phenomena trigger questions, so I usually start the lesson with a question-and-answer session. (Mita)
>
> The second method is assigning tasks based on phenomena or images, often using graphs. The interpretation of these graphs by students often reveals whether they still misunderstand concepts or have correctly grasped them. (Mita)
>
> Next, during the main learning activities, I begin by providing student worksheets. During this time, discussions usually occur, and students tend to openly express their true understanding. (Mita)

The data obtained indicates that using tests to detect physics misconceptions is generally not preferred by all teachers. Instead, the methods commonly chosen are in the form of interviews for several reasons already confirmed in previous sections.

### 3.3 The Examples of Learning Material

The interviews revealed several topics frequently mentioned by participants, whether Physics teachers in secondary schools or higher education. Some supporting statements include Budi's remark "…because there are still students who cannot differentiate between distance and displacement." This aligns with other participants' statements:

> When asked, even higher education students [Physics or Physics education majors] respond to Newton's Second Law with F = m.a or Newton's Third Law with the action-reaction principle. Their answers are often correct, but they do not understand the physical meaning of action and reaction (Baim).
>
> There is also a misconception related to the concept of force. They think that a larger force means faster acceleration (Mita).
>
> Many students are still diagnosed with Physics misconceptions, particularly in distinguishing between weight and mass (Mita).
>
> For example, in the market, they [students] consider weight to be mass. And when they mention distance, they just say "kilo" not specifying kilometers (Aswar).

Baim frequently mentioned his views on students in several parts of his statements. Baim is a Physics teacher at a secondary school who often mentor's student teachers during their teaching practice. Based on Baim's statements regarding **Newton's Laws**, Mita's statements on **Force and Motion**, and Quantities and Units, and Aswar's statements on **Quantities and Units**, it indicates that these topics often lead to student misconceptions. These topics in Physics are categorized under **Mechanics** or study about motion.

Some teaching experiences shared by Aswar and their impact on his view of misconceptions are illustrated in the following quote

> One memorable experience about two years ago was when I was teaching the concept of force and motion in Grade X. There was a student named Budi who was very convinced that heavier objects always fall faster than lighter ones, despite my explanations and demonstrations. Budi remained steadfast in his belief. Such experiences greatly influenced my perspective as a teacher. I recognize that changing misconceptions is not always straightforward and often requires a personalized approach. (Aswar)

From the above quotes, it is evident that misconceptions are prevalent across various educational levels, including secondary and higher education, particularly in the topics confirmed by the participants.

**4. Teachers' reflections on future assessment practice (RQ3)**

Teacher reflections for future assessments were explored by asking several key questions, including reflections on the assessments they have conducted so far and their future. The data obtained is presented in Table 3.

Table 3. Teachers' Reflection for Future Assessments

| No. | Category | Teacher's Reflection |
|---|---|---|
| 1. | Test development | - Create separate tests without combining them with formative tests<br>- Create a medium-powered test that looks appealing<br>- Develop tests according to the character of students |
| 2. | Validity and reliability of test | - Using tests developed by ready-to-use researchers<br>- Using standardized tests in the Physics Education community |
| 3. | Diagnostic tools | - Question redaction in narrative form (without nominal)<br>- Essay<br>- Multiple-choice test<br>- Developing interview methods |

The information in Table 3 reveals that participants responded with their plans for future assessments. Generally, five points emerged: the development of misconception tests should not be combined with other tests like formative assessments; some teachers expressed willingness to use reliable instruments in their teaching; test presentation should utilize technological advancements for ease; adjustments to test formats should consider student characteristics and preferred methods. Some teachers plan to focus on test creation, while others decided to continue using interviews, considering them the most effective and practical method given their busy schedules. Some participant responses include:

> ...Teachers must keep up with the times. In the past, teachers rarely used technology, perhaps it was not yet the era. I frequently use Quizziz because students enjoy seeing their test results immediately (Hardi).

> ...after our discussion, it seems we need a separate instrument. With a specific instrument, we can truly understand how well students grasp a basic competence and appropriately follow up (Dodi).

From the above explanations, it is evident that Hardi has reflected on previous learning periods. How his students, initially passive, responded positively through technology-assisted applications. Dodi felt it is crucial to provide specific instruments for diagnostic tests to detect student misconceptions. Efforts to optimize include designing the learning process to be engaging, complemented by numerous discussion sessions. The goal is to improve students' understanding of Physics concepts compared to previous periods. This is confirmed in the statements of Mita and Aswar:

> The first step is how to reactivate question and answer sessions in every interaction or learning process. This Q&A can then serve as the initial step to identify whether students have not yet understood the concepts. Secondly, administering diagnostic tests. In the core activities, I will reactivate group discussions." (Mita)

> Since then, I have begun developing more interactive and experimental learning strategies. This experience has taught me the importance of patience, creativity in teaching methods, and the value of experiential learning. (Aswar)

Reflections on the diagnostic test planned and their views on assessments are represented by statements like Dani's: "...multiple-choice or matching tests could also be used, depending on the level of the Physics material." This is further supported by statements from Dodi and Budi

> ...if I want to assess misconceptions, I will create essay questions and minimize the mathematics content, focusing more on open-ended questions. (Dodi)

> ...in my view, it's about returning assessments to their true essence. If we consider the assessment pyramid, especially assessment as learning, assessments should serve as a facility for students to improve their concepts. (Budi)

The reliability of the tests to be used is a key consideration for participants. Regarding the reliability of the diagnostic tests, when Baim and Budi were asked about implementing standardized tests to detect students' Physics misconceptions, they confirmed

> ...It would be excellent because, as teachers or lecturers, we often create questions without conducting research. If the test's reliability is proven, it would be very beneficial. (Baim)

> ...if there are experts specializing in research for misconception assessments that are ready-to-use and tested, it would greatly assist our teachers. (Budi)

Participants' views on their hopes or plans for conducting diagnostic tests for Physics misconceptions in the future are shown in statements from Nana and Aswar:

> ...if it suits the character of my students, why not use it in school? But if the instrument doesn't align with my students' characteristics, it might not be used." (Nana)

> ...in my opinion, if what students read is not interesting, their answers will be haphazard. They tend to get bored and provide answers that don't align with what we want to measure. That's just my personal view." (Aswar)

From the above explanations, it is evident that participants, in choosing and considering methods to assess or detect students' Physics misconceptions, consider not only practicality for use in the learning process but also the characteristics of the students they face in their respective teaching locations.

## Discussion

While most teachers are aware of the existence of misconceptions, they do not yet understand how to develop their roles in teaching (Gomez-Zwiep, 2008). Another fact confirmed by Burgoon et al., (2011) shows that there are similarities in misconceptions experienced by both teachers and students on certain topics, and Moodley & Gaigher (2019) Moodley (2019) found that teachers' understanding of misconceptions is influenced by their educational background. The study of misconceptions is a continuously researched topic within the physics education research (PER) community (Docktor & Mestre, 2014). Over the past few decades, numerous studies on Physics misconceptions have been reported. Most of these studies highlight the need to focus on professional development for teachers, particularly in reducing student misconceptions. In our study, we used a phenomenological approach to explore teachers' perceptions of Physics misconceptions, their knowledge of diagnostic tools that can be used, their experiences in diagnosing misconceptions, and their reflections on these experiences. Our discussion is based on these descriptions.

The personal experiences of Physics teachers were explored through themes such as teacher motivation, teachers' perceptions of their profession, and teachers' knowledge of student characteristics. From this section, we discovered how participants' roles as Physics teachers connect to their personal experiences. Regarding motivation, participants generally had diverse reasons for choosing Physics teaching as a profession. The data showed that teachers' views on their profession combine their experiences as students and as educators. Teachers' knowledge of their students' characteristics is understood in various contexts, including material presentation, assessment evaluation, gender, and interests.

These findings align with Kyriacou & Coulthard (2018), who identified three fundamental reasons individuals choose the teaching profession. First, altruistic reasons, viewing teaching as beneficial to social life, are seen in data such as (1) inspiration from Physics teachers remembered for their warm interactions with students; (2) how participants position themselves as facilitators, parents, or friends. Second, intrinsic reasons encompass the activity of teaching children and a person's interest in utilizing their knowledge. This is evidenced by how participants understand their students' characteristics in various contexts. Third, extrinsic reasons include salary, social status, etc. One extrinsic reason noted in the data is that the teaching profession is viewed as the quickest route to employment.

Participants' perceptions of Physics misconceptions view them as a natural occurrence among students. According to the participants, students enter the classroom with prior knowledge and intuitions based on their everyday experiences, which sometimes conflict with actual Physics principles, making it the teacher's responsibility to correct these misconceptions. The data indicates that participants are aware of the potential impact that Physics misconceptions can have on students. The recurrence of certain topics that often lead to misconceptions (in this study, typically occurring in mechanics) highlights the need for attention to students' Physics misconceptions in both secondary and higher education. Misconceptions in secondary school can persist into higher education and those in higher education may continue when students become Physics teachers (Suprapto, 2020). This is further reinforced by the data indicating that similar topics are conducive to the emergence of Physics misconceptions among students at both educational levels (middle school and higher education).

Regarding teachers' knowledge of diagnostic tools for Physics misconceptions, it appears that this knowledge is not widely known among participants, particularly for Physics teachers at middle schools. The data indicates that training provided to participants related to assessment (including Physics misconception assessment) has not been conducted for Physics teachers at middle schools. The training they have received generally involves socialization during curriculum changes. Physics instructors in higher education confirmed that they have conducted several training sessions for specific teacher communities on assessing Physics misconceptions, but these efforts have only reached certain locations. In this context, strengthening communities/institutions as learning platforms for participants plays a crucial role.

So far, the practice most chosen by participants for diagnosing Physics misconceptions is interviews. The considerations for choosing interviews include (1) not requiring much technical preparation; (2) providing quick feedback; (3) concerns that answers in diagnostic tests may result from student collaboration. Several opinions on interviews view them as a detailed method for detecting student misconceptions (Brookhart & Nitko, 2019). Other supportive opinions state that students' ideas and conceptual understanding structures can be revealed through interviews (Fisher & Frey, 2014; Mintzes et al., 2005). Due to these advantages, Hestenes & Wells (1992) stated that interviews could confirm students' responses in identifying misconceptions. However, several drawbacks of interviews should be considered, including (1) they are time-consuming, especially with large samples; (2) they require interview skills from teachers, including building trust with students for in-depth interviews; (3) there is a tendency for subjective assessment of interview results (Kaltakci Gurel et al., 2015). Another method is to evaluate from the results of formative assessments (with diagnostic misconceptions integrated into formative tests).

The selection of interviews as a practical method by participants or the decision to integrate diagnostic misconceptions into daily tasks or formative assessments, based on the data we obtained, leads to the argument that this also relates to the prevailing curriculum. Participants with over 10 years of teaching experience confirmed that when the curriculum still included National Exams and the extensive learning material requirements that had to be completed within one academic year, several participants focused solely on preparing their students for the national exams. These exams generally involved extensive calculations, leading participants to design learning strategies that provided tips and tricks for recognizing problem types and solving them. In this context, conceptual understanding was not prioritized as the emphasis was on mastering mathematics to solve exam problems. Participants confirmed that the new curriculum no longer includes as much material as previous curricula. Those teaching in secondary schools feel they are given more time to optimize material presentation. The new curriculum allows participants to prioritize conceptual understanding in their teaching, with the hope that misconceptions can be addressed appropriately.

Several teaching reflections were confirmed in this study, considering various aspects by the participants. These include test development, test validity and reliability, and the test format (multiple-choice, essay/open-ended, and interviews). Multiple-choice tests as diagnostic tools have their advantages and disadvantages. Time efficiency for answer analysis and more objective assessment are the strengths of MCT (Brookhart & Nitko, 2019; Kaltakci Gurel et al., 2015; Mintzes et al., 2005). Even with many participants, the time required to analyze answers is shorter compared to open-ended tests. However, some weaknesses of MCT include the inability to investigate students' ideas in-depth, the possibility of correct answers by guessing, and the difficulty in constructing appropriate items (Brookhart & Nitko, 2019; Kaltakci Gurel et al., 2015; Mintzes et al., 2005). The limited answer choices restrict the exploration of new misconception patterns.

Open-ended tests have the advantage of revealing unexpected errors or misunderstandings (Mintzes et al., 2005). These tests allow students to express their conceptual understanding in their own words, possibly providing unexpected responses (Kaltakci Gurel et al., 2015). However, the drawbacks include the lengthy time required to analyze student responses and the difficulty in assessing and categorizing them (Kaltakci

Gurel et al., 2015). These reasons make open-ended tests less suitable for studies with large participant numbers.

### Overall Essence of the Experience

The overall reporting of data and discussions presented leads us to the statement about the essence of the experiences of all participants involved in our study. Each participant strives to optimally present material to help students understand concepts. Their diverse personal experiences and motivations for becoming Physics teachers drive them to understand their students' characteristics in various contexts. Every participant acknowledges the occurrence of Physics misconceptions in their classes. Interviews were chosen by all eight participants as the method for diagnosing Physics misconceptions. Interviews are considered the easiest way to detect and provide immediate feedback if students experience misconceptions. The eight participants also openly agreed to reflect on and improve their practices in assessing Physics misconceptions in the future.

## Conclusion

The investigation conducted with eight Physics teachers provided several insights into their perceptions of diagnostic tests for student misconceptions in Physics. The findings indicate that theoretical understanding of diagnostic tests for Physics misconceptions is generally not widely known among Physics teachers. Participants view Physics misconceptions as a natural occurrence among students, which can arise from various conditions and have long-term impacts.

Furthermore, the application of Physics misconception assessments is linked to the prevailing curriculum. The existence of National Exams and the extensive material requirements within one academic year caused participants to prioritize exam preparation over addressing misconceptions. Currently, the implementation of Physics misconception assessments by participants is mainly through interviews/direct questioning and formative tests/quizzes. In other words, teachers' plans for implementing Physics misconception assessments in the future include developing reliable standalone misconception tests tailored to teachers' needs.

### Implications

The phenomena described in this study highlights the necessity of paying attention to the community of Physics teachers regarding the assessment of student misconceptions. The study's findings indicate the need for facilities from relevant agencies, such as those in the Education sector, to provide essential training related to this matter. Additionally, the curriculum significantly impacts how teachers present Physics lessons in the classroom and the assessment methods they choose. Nowadays, Physics, as a part of STEM, has relatively few enthusiasts worldwide, making awareness of the importance of teachers' roles in increasing student interest essential. The study's results are expected to serve as a reference for the government, teachers, and preservice teachers.

## Author Note

Mutmainna is a lecturer of the Physics Education Study Program, University of West Sulawesi, Indonesia. He was appointed as a lecturer in 2019 and continued his studies at the Doctoral Program in Education Research and Evaluation, Yogyakarta State University, Indonesia. His research interests include Physics education, learning assessment and evaluation, and curriculum. She can be reached by email: mutmainna.2022@uny.ac.id or mutmainna_kadir@unsulbar.ac.id

Edi Istiyoono has been a lecturer in the Ph.D. program in Educational Research and Evaluation at Yogyakarta State University since 2015. The second author’s expertise located in Measurement, Assessment, and Evaluation of Physics Education. Several of the second author’s research projects align with the research topic, particularly concerning diagnostic tests for physics misconceptions, which have been explored in recent years. He can be reached by email: edi_istiyono@uny.ac.id

Harianto has been a lecturer in the Ph.D. program in Educational Research and Evaluation at Yogyakarta State University since 2010. The third author’s expertise is in Research and Evaluation of Electrical Engineering Education. The third author teaches courses relevant to the research topic and supervises Ph.D. students in the measurement concentration. He can be reached by email: haryanto@uny.ac.id

Heri Retnawati has been a lecturer in the Ph.D. program in Educational Research and Evaluation at Yogyakarta State University since 2009. The fourth author's expertise is in Mathematics Education Assessment. The fourth author teaches courses relevant to the research topic and supervises Ph.D. students in the measurement concentration. She can be reached by email: heri_retnawati@uny.ac.id

Caly Setiawan has been a lecturer in the Ph.D. program in Educational Research and Evaluation at Yogyakarta State University since 2017. The fifth author's expertise is in Physical Education and Sports Pedagogy. The fifth author teaches courses relevant to the research topic, particularly in Qualitative Research Methods and Qualitative Data Analysis Techniques. He can be reached by email: csetiawan@uny.ac.id

## Acknowledgement

The first author would like to thank the Ministry of Education, Culture, Research, and Technology (KEMENDIKBUDRISTEK), the Center for Higher Education Funding (BPPT), Education Fund Management Institution (LPDP) of the Republic of Indonesia, and Indonesian Education Scholarship (BPI) for funding in the first author's study doctoral and this research.